\documentclass[10pt,superscriptaddress,tightenlines,twocolumn,amsmath,amssymb,aps, pra]{revtex4-2}
\usepackage{amsfonts}
\usepackage{amsmath}
\usepackage{amsthm}
\usepackage{amscd}
\usepackage{amssymb}
\usepackage{subfigure}
\usepackage{amsxtra}
\usepackage{gensymb}
\usepackage{bm}           
\usepackage{bbm}
\usepackage{graphicx}
\usepackage{epstopdf}
\usepackage{color}
\usepackage{hyperref}
\usepackage{makecell}
\usepackage{multirow}
\usepackage{lettrine}

\hypersetup{colorlinks=true, linkcolor=blue, citecolor=red, urlcolor=blue}

\def\bi#1\ei {\begin{itemize}#1\end{itemize}}
\def\bn#1\en {\begin{enumerate}#1\end{enumerate}}
\def\bea#1\eea {\begin{align}#1\end{align}}
\def\bean#1\eean {\begin{align*}#1\end{align*}}
\def\ben#1\een {\begin{equation*}#1\end{equation*}}
\def\be#1\ee {\begin{equation}#1\end{equation}}
\def\bes#1\ees {\begin{equation}\begin{split}#1\end{split}\end{equation}}
\def\bear#1\eear {\begin{eqnarray}#1\end{eqnarray}}
\def\bear#1\eear {\begin{eqnarray*}#1\end{eqnarray*}}

\newcommand{\beq}{\begin{equation}}
\newcommand{\eeq}{\end{equation}}

\begin{document}

\title{Experimental quantum advantage with quantum coupon collector}
\author{Min-Gang Zhou}\thanks{These authors contributed equally to this work}
\author{Xiao-Yu Cao}\thanks{These authors contributed equally to this work}
\author{Yu-Shuo Lu}\thanks{These authors contributed equally to this work}
\author{Yang Wang}
\author{Yu Bao}
\author{Zhao-Ying Jia}
\affiliation{National Laboratory of Solid State Microstructures, School of Physics, Collaborative Innovation Center of Advanced Microstructures, Nanjing University, Nanjing 210093, China}
\author{Yao Fu}
\affiliation{MatricTime Digital Technology Co. Ltd., Nanjing 211899, China}	
\author{Hua-Lei Yin}\email{hlyin@nju.edu.cn}
\affiliation{National Laboratory of Solid State Microstructures, School of Physics, Collaborative Innovation Center of Advanced Microstructures, Nanjing University, Nanjing 210093, China}
\author{Zeng-Bing Chen}\email{zbchen@nju.edu.cn}
\affiliation{National Laboratory of Solid State Microstructures, School of Physics, Collaborative Innovation Center of Advanced Microstructures, Nanjing University, Nanjing 210093, China}
\affiliation{MatricTime Digital Technology Co. Ltd., Nanjing 211899, China}

\date{\today}

\begin{abstract}
An increasing number of communication and computational schemes with quantum advantages have recently been proposed, which implies that quantum technology has fertile application prospects. However, demonstrating these schemes experimentally continues to be a central challenge because of the difficulty in preparing high-dimensional states or highly entangled states. In this study, we introduce and analyze a quantum coupon collector protocol by employing coherent states and simple linear optical elements, which was successfully demonstrated using realistic experimental equipment. We showed that our protocol can significantly reduce the number of samples needed to learn a specific set compared with the classical limit of the coupon collector problem. We also discuss the potential values and expansions of the quantum coupon collector by constructing a quantum blind box game. The information transmitted by the proposed game also broke the classical limit. These results strongly prove the advantages of quantum mechanics in machine learning and communication complexity. 
\end{abstract}

\maketitle

\bigskip
\noindent
\textbf{Introduction}

The ‘second quantum revolution’ aims to explore the superiority of quantum resources over classical resources in terms of communication, computation, and artificial intelligence. To demonstrate that this goal is feasible in practice, a series of schemes with quantum advantages were experimentally implemented. These schemes included improving on the security of communication \cite{pappa2014experimental,xu2020secure,PRXQuantum.2.040334,Fu2015Long,Zhang2017Quantum,lucamarini2018overcoming,Gu:21,alikhani2021experimental,li2021finite,sheng2022one,long2021drastic,qi202115}, enhancing computational power for specific tasks \cite{neville2017classical,gong2021quantum,zhong2020quantum,broome2013photonic,bentivegna2015experimental,zhong2019experimental,farhi2016quantum,bremner2017achieving,bravyi2018quantum,gao2017quantum,bermejo2018architectures,boixo2018characterizing,arute2019quantum} and reducing the necessary resources used to complete specific communication tasks \cite{horn2005single,xu2015experimental,guan2016observation,kumar2019experimental,arrazola2018quantum,centrone2021experimental}.
In addition, machine learning can extract useful knowledge from data, which can then have a significant impact on productivity, technology, and the economy \cite{goodfellow2016deep}. This has led to an increasing interest in the question of quantum machine learning \cite{biamonte2017quantum}: Can we improve machine learning by using quantum resources? Owing to the unique entanglement properties of quantum states, quantum models may be able to produce atypical patterns that cannot be effectively produced by classical models or effectively reduce training time. Therefore, some studies have made bold attempts. For example, in Ref. \cite{bondarenko2020quantum}, a quantum autoencoder that can successfully denoise specific quantum states subjected to specific noises was developed. Reference \cite{cong2019quantum} described a quantum neural network that can accurately recognize quantum states associated with a one-dimensional symmetry-protected topological phase. Moreover, there are several other excellent studies \cite{beer2020training,chen2018quantum,wan2017quantum,killoran2019continuous,torrontegui2019unitary,arrazola2019machine}.

On the other hand, most of these attempts are heuristic and have not theoretically proven that quantum machine learning exhibits a better performance or shorter training time than classical machine learning. The training time of a model includes the time complexity of the learning algorithm. If the algorithm takes a constant amount of time for processing each sample, the concern for the time complexity translates into a concern of the sample complexity. Probably approximately correct (PAC) learning theory \cite{valiant1984theory,haussler1990probably} provides the minimum number of samples necessary for a learning algorithm to complete a learning task. Researching quantum machine learning using this theory can therefore lay a positive theoretical foundation for exploring quantum advantages in machine learning.

For the first time, Ref.~\cite{arunachalam2020quantum} described the use of PAC learning theory to rigorously prove that quantum technology can provide a learning algorithm with quantum advantage for machine learning. The method applied clever quantum measurements to the learning task known as the `coupon collector problem' \cite{motwani2010randomized}. Specifically, Ref. \cite{arunachalam2020quantum} gives a surprising result: for the coupon collector problem, the sample complexity of the quantum learning algorithm does not change with changes in the search space of the algorithm. This result is impossible in classical machine learning \cite{valiant1984theory}. However, the experimental demonstration of this algorithm~\cite{arunachalam2020quantum} requires high-dimensional states that are difficult to prepare. Even if these states are decomposed into a tensor product of qubits, these qubits must be highly entangled \cite{horn2005single, du2006experimental, de2004one}. These requirements are far beyond the scope of current technology. More seriously, the algorithm requires a specific projective measurements, which is difficult to implement in experiment.

In this work, we experimentally demonstrate the quantum coupon collector algorithm by proposing a coherent-state quantum coupon collector protocol. Our protocol avoids the abovementioned difficulties. To do this, our protocol not only maintains the important properties of the original one, but also introduces new conceptual tools that can be implemented using only linear optics operations and single-photon detectors. Even without a quantum computer, these tools enable us to demonstrate the quantum advantages in machine learning at the current technological level. Moreover, the coupon collector problem can be considered as a communication task. Similar to quantum fingerprinting \cite{xu2015experimental,guan2016observation}, our protocol also experimentally demonstrates the advantages of quantum mechanics in the context of communication complexity. These results, in addition to their fundamental interest \cite{brassard2003quantum,steane2000physicists,buhrman2010nonlocality}, will further inspire new designs of communication systems, large-scale integration circuit designs, and data structures \cite{kushilevitz1997communication}, thus paving the way for other communication or computational tasks that rely on similar principles.

\bigskip
\noindent
\textbf{Results}

\noindent
\textbf{Coherent-state quantum coupon collector protocol.} 

To clearly describe our protocol, we briefly introduce the coupon collector problem \cite{motwani2010randomized}. This problem can be abstracted as learning exactly an unknown set $S$. Specifically, this set $S$ is limited to a subset of the set $[n]:=\{1, \ldots, n\}$, where the size $k~(k < n)$ of set $S$ is known. To learn the set $S$, several copies of $S$ are given, and only one element is allowed to be extracted from each copy. The task is to determine the minimum number of copies required to learn $S$ exactly.

Because the elements in $S$ are independent of each other, the best strategy for learning $S$ is to randomly extract one of these elements in each copy. Under this strategy, if $i~(i<k)$ distinct elements have been obtained, the expected number of copies needed to learn the $(i+1)$th element is $k/(k-i)$. Therefore, the expected number of copies needed to learn $S$ is $\sum_{i=0}^{k-1} \frac{k}{k-i} \sim k \ln k$. Continuing based on this, Ref. \cite{motwani2010randomized} shows that $\Theta(k \log_2 k)$ copies are necessary and sufficient for learning $S$ with a high probability. 


However, if these copies are quantum copies in the form of $|S\rangle:=\frac{1}{\sqrt{k}} \sum_{i \in S}|i\rangle$, the number of copies required can be further reduced. This is because we can perform more quantum operations on $|S\rangle$ before measuring them on the computational basis. In other words, by using quantum copies $|S\rangle$, the strategy for learning $S$ is no longer limited to random sampling. Reference~\cite{arunachalam2020quantum} shows that when the number of missing elements $m=n-k$ is very small, the number of copies of $|S\rangle$ used to learn $S$ is reduced to $\Theta(k \log_2 (m+1))$. The method first performs a $2$-outcome projective measurement with operators $|[n]\rangle\langle[n]|$ and $I- |[n]\rangle\langle[n]|$ on copies of $|S\rangle$, where $|[n]\rangle:=\frac{1}{\sqrt{n}} \sum_{i \in[n]}|i\rangle$. After the measurement, the second outcome $|\psi\rangle=\sqrt{\frac{m}{n}}|S\rangle-\sqrt{\frac{k}{n}}|\bar{S}\rangle$ is obtained with probability $m/n$, where $\bar{S}$ is the set of missing elements, and $|\bar{S}\rangle:=\frac{1}{\sqrt{m}} \sum_{i \in \bar{S}}|i\rangle$. Then, $|\psi\rangle$ is measured on the computational basis to obtain the missing elements. This method transforms $|S\rangle$ into $|\psi\rangle$, and then infers the elements in $|S\rangle$ by learning the elements in $|\bar{S}\rangle$ from $|\psi\rangle$. Because $|\bar{S}\rangle$ has fewer elements, this method reduces the number of copies required to learn $S$.

However, difficulties arise when we attempt to demonstrate this method experimentally. This is because it is difficult to experimentally construct a single $k$-dimensional quantum state \cite{massar2005quantum, garcia2013swap, arrazola2014quantum}. Even if this state is decomposed into a tensor product of qubits, these qubits must be highly entangled \cite{horn2005single, du2006experimental, de2004one}. This is also not feasible using current technology. More seriously, the operators $|[n]\rangle\langle[n]|$ and $I- |[n]\rangle\langle[n]|$ in this method are difficult to implement in experiments.

\begin{figure*}
\includegraphics[width=18cm]{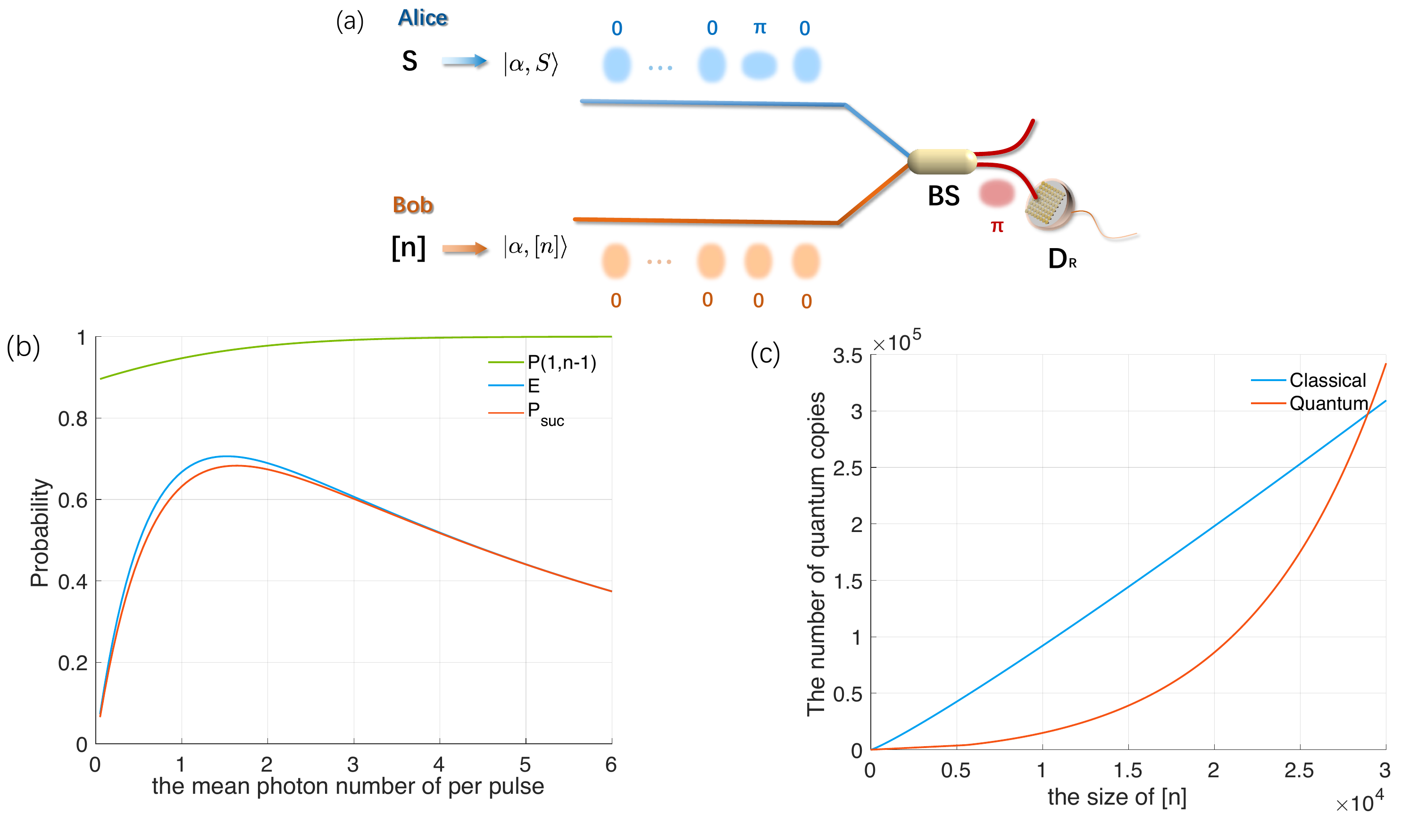}
\caption{(a) Coherent-state quantum coupon collector protocol. Alice and Bob generate pulses $|\alpha,S\rangle$ and $|\alpha,[n]\rangle$ according to $S$ and $[n]$, respectively. The pulses travel along their respective routes to interfere at the beam splitter (BS) and are then detected by a single-photon detector $D_R$. After the experiment, Bob learns the elements in set $S$ according to the results of $D_R$. (b) Efficiency $E$, correct probability $P(1,n-1)$ and success probability $P_{\rm suc}$ versus different mean photon numbers per pulse. We draw the numerical results under $n=4000$, $k=n-1$, $p_d=10^{-8}$, $\nu=0.99998$, and $\eta=68\%$. (c) The required samples for the classical and quantum protocols versus the size of the set $[n]$ with $p_d=10^{-8}$, $\nu=0.99998$, and $\eta=68\%$.}
\label{protocol}
\end{figure*}

Therefore, we introduce an alternative scheme, which is defined as the `coherent-state quantum coupon collector protocol'. This scheme maintains the main idea of the original one, which is to learn $S$ by measuring the missing elements. In addition, this scheme uses a sequence of coherent states to implement copies of $|S\rangle$. Coherent states are easy to prepare and can be transformed using simple linear optical elements. Therefore, this scheme is particularly attractive from a practical point of view.

In our scheme, copies of $|S\rangle$ are implemented using a time sequence of $n$ weak coherent optical pulses
\beq
|\alpha,S\rangle:=\bigotimes_{i=1}^{n}\left|(-1)^{j} \alpha\right\rangle_{i},
\eeq where $\alpha$ is a complex number, and $\left|(-1)^{j} \alpha\right\rangle_{i}$ is a coherent state with amplitude $\alpha$ at the $i$th time mode, where $j=0$ for $i \in S$ and $j=1$ otherwise. The phases of these coherent pulses depend on $S$, but their intensities are the same. Thus, the state $|\alpha,S\rangle$ has the mean photon number $\mu=n|\alpha|^{2}$.
Note that $|S\rangle$ is given by projecting the state $|\bar{\alpha},S\rangle:=\bigotimes_{i \in S}\left| \alpha\right\rangle_{i}$ into the single-photon subspace. The total intensity $k|\alpha|^2$ of $|\bar{\alpha},S\rangle$ represents the number of copies of $|S\rangle$. However, our scheme uses $|\alpha,S\rangle$ instead of $|\bar{\alpha},S\rangle$. In other words, when $i \notin S$, our scheme sends state $\left|- \alpha\right\rangle_i$ instead of a vacuum state. This method can improve the efficiency of detecting missing elements, but it also causes the number of copies by our scheme to be slightly different from those of a scheme using $|\bar{\alpha},S\rangle$. Specifically, the number of copies by our scheme are at most $O(n|\alpha|^2)$. For comparison, the number of copies by a scheme using $|\bar{\alpha},S\rangle$ are at most $O(k|\alpha|^2)$. However, as we will see later, this subtle difference can be ignored, and using $|\alpha,S\rangle$ greatly improves the efficiency of detecting missing elements.

As we discussed previously, our scheme maintains the important characteristics of the original one, that is, the time bin $i$ of state $\left|- \alpha\right\rangle_{i}$ in $|\alpha,S\rangle$ is found through complementary measurements, and the elements in $S$ are derived from these time bins. To do this, a local state $|\alpha,[n]\rangle$ is prepared and sent to a $50:50$ beam splitter (BS) to interfere with $|\alpha,S\rangle$ (Fig.~\ref{protocol}(a)), where

\beq
|\alpha,[n]\rangle:=\bigotimes_{i=1}^{n}\left|\alpha\right\rangle_{i}.
\eeq The interference result is recorded by a single photon detector $D_R$. If the detector $D_R$ clicks at the $i$th time bin, then we consider the pulse of the $i$th time bin in $|\alpha,S\rangle$ to be $\left|- \alpha\right\rangle_{i}$. Thus, all time bins containing $\left|- \alpha\right\rangle$ in $|\alpha,S\rangle$ can be learned exactly from the outcomes of the detector $D_R$. Without loss of generality, let Alice be a coupon maker and Bob be a coupon collector. Bob's task is to learn all elements in $S$ from the state $\left|\alpha,S\right\rangle$ prepared by Alice. The detailed steps are described as follows.

(1) Alice selects $k$ elements from set $[n]$ as set $S$ and selects an appropriate value $|\alpha|^2$ as the intensity of each pulse.

(2) Alice encodes the pulses $|\alpha,S\rangle$ according to $S$ and the intensity $|\alpha|^2$; if $i \in S$, Alice prepares a coherent pulse $\left|\alpha\right\rangle_{i}$ at the $i$th time bin and sends it to Bob; otherwise, Alice sends $\left|-\alpha\right\rangle_{i}$ to Bob.

(3) Alice announces the value of $|\alpha|^2$, the elements of $[n]$, and the size $k$ of $S$.

(4) Bob encodes the pulses $|\alpha,[n]\rangle$ according to $[n]$ and intensity $|\alpha|^2$: Bob prepares a coherent pulse $|\alpha\rangle$ for all time bins.

(5) Bob uses a $50:50$ beam splitter and two single-photon detectors to perform interference measurements on the pulses $|\alpha,S\rangle$ and $|\alpha,[n]\rangle$.

(6) Bob records whether the detector $D_R$ clicks or not at each time bin, and then learns $S$ exactly based on the outcomes of the detector $D_R$ and the size $k$ of the set $S$.

At each time bin $i$, the output state after the BS is $\left|\frac{\left((-1)^{j}+1\right) \alpha}{\sqrt{2}}\right\rangle_{D_L, i} \otimes\left|\frac{\left((-1)^{j}-1\right) \alpha}{\sqrt{2}}\right\rangle_{D_R, i}$. Because $D_L$ is unnecessary in this work, it is not drawn in Fig.~\ref{protocol}(a). It is easy to verify that in the ideal case, the value of $j$ determines whether the pulses output by the BS go to $D_L$ or $D_R$, thus helping Bob to learn the state $|\alpha,S\rangle$.

However, even under ideal conditions, the proposed scheme has an intrinsic failure probability. This is because a coherent pulse may collapse to a vacuum state after being measured, and thus, cannot be detected by detector $D_R$. Specifically, when Alice sends the coherent pulse $\left|\alpha\right\rangle_{i}$, $D_R$ will never click. Therefore, Bob can always obtain the correct results. However, when Alice sends a coherent pulse $\left|-\alpha\right\rangle_{i}$, the detector $D_R$ has a non-click probability of $P_{\rm not}=e^{-2|\alpha|^2}$ (the detection efficiency is assumed to be $100\%$ in the ideal case). Note that if Alice sends a vacuum state when $i \notin S$, then the non-click probability of $D_R$ is increased to $e^{-\frac{|\alpha|^2}{2}}$. This is why when $i\notin S$, our scheme sends $\left|- \alpha\right\rangle_i$ instead of a vacuum state.
Only when detector $D_R$ detects all $\left|-\alpha\right\rangle_{i}$ sent by Alice can we infer the elements in $S$. Therefore, the success probability of our scheme without experimental imperfections is $P(m)=(1-P_{\rm not})^m$, where $m$ is the number of missing elements of $S$.
Note that although the use of the coherent state sequence $|\alpha,S\rangle$ for implementing copies of $|S\rangle$ is easier to demonstrate experimentally, it also introduces an intrinsic failure probability. Fortunately, as we will see later, this failure probability is negligible compared to the failure probability introduced by experimental imperfections.

To demonstrate the quantum advantage of our scheme experimentally, we need to eliminate the influence of the failure probability as much as possible. On the one hand, we can reduce the failure probability by increasing the mean photon number per coherent pulse. On the other hand, we can calculate the required number of copies based on the expectation of $100\%$ success. However, these methods increase the number of copies required. Therefore, the selection of an appropriate mean photon number is particularly important.

\bigskip
\noindent
\textbf{Protocol in the presence of experimental imperfections.} 

So far, we have only discussed the success probability of our protocol under ideal conditions. However, owing to experimental imperfections, the success probability formula of our protocol must be modified. We consider imperfect experimental models characterized by three parameters: the combined effects of limited detector efficiency and channel loss $\eta$, dark count rate $p_d$ of the single-photon detector, and the limited visibility $\nu$ of the interferometer.

By replacing $|\alpha,S\rangle$ with $|\sqrt\eta \alpha,S\rangle$, we can eliminate the effect of $\eta$ without changing the form of the success probability formula. However, $p_d$ and $\nu$ will cause $D_R$ to click at incorrect time bins with a non-zero probability; thus, the success probability formula must be modified.
In this case, when Alice sends a pulse $\left|\alpha\right\rangle_{i}$, the probability that detector $D_R$ clicks is given by
\beq
P_{\alpha}=\left(1-e^{-2(1-v) |\alpha|^2\eta}\right)+p_{\mathrm{d}},
\eeq and when Alice sends a pulse $\left|-\alpha\right\rangle_{i}$, the probability that the detector $D_R$ clicks is
\beq
P_{-\alpha}=\left(1-e^{-2|\alpha|^2\eta}\right)+p_{\mathrm{d}}.
\eeq
Because $D_R$ may click when Alice sends $\left|\alpha\right\rangle_{i}$ and may not click when Alice sends $\left|-\alpha\right\rangle_{i}$, the number of clicks of the detector $D_R$ may be different from the size $k$ of $S$. In this case, Bob can determine that the experimental results are not available and must be discarded. Therefore, Alice and Bob need to repeat multiple experiments to obtain usable experimental data. Here, we define a new efficiency $E=M/N$ to measure the ratio of the number $M$ of available experimental results of the total number $N$ of experiments, which can be calculated by the following formula:

\beq
E = \sum_{i=0}^{m}D^{mi}_{-\alpha}D^{k(m-i)}_{\alpha}
\eeq where $D^{mi}_{-\alpha}=C_{m}^{i}P_{-\alpha}^{i}(1-P_{-\alpha})^{m-i}$ represents the sum of the probabilities of all possible cases where $D_R$ detects $i$ out of $m$ coherent states $\left|-\alpha\right\rangle$ sent by Alice, and $D^{k(m-i)}_{\alpha}=C_{k}^{m-i}P_{\alpha}^{m-i}(1-P_{\alpha})^{k-m+i}$ represents the sum of the probabilities of all possible cases where $D_R$ detects $m-i$ out of $k$ coherent states $\left|\alpha\right\rangle$ sent by Alice.

In addition, even if Bob obtains usable experimental results, $D_R$ has a probability of clicking at the wrong time bin, thus making him misjudge the elements in $S$. This requires us to define the correct probability $P(m,k)=P_{-\alpha}^{m}(1-P_{\alpha})^k/E$ when the experimental results are available. Thus, the success probability is modified to $P_{\rm suc} = E \times P(m,k)$. Based on the expectation of $100\%$ success, the number of quantum samples required to complete the task is
\beq
R = \frac{n|\alpha|^2}{P_{\rm suc}}.
\eeq
Without loss of generality, we choose a special set $S_j$ to verify the quantum advantage of our protocol, where $S_j=\{1,2,...,j-1,j+1,...,n\}$ and size $k_j := |S_j| = n - 1$. Therefore, the correct probability can be simplified as
\beq
P(1,n-1)=\frac{P_{-\alpha}(1-P_\alpha)^{n-1}}{D^{10}_{-\alpha}D^{(n-1)1}_{\alpha}+D^{11}_{-\alpha}D^{(n-1)0}_{\alpha}}.
\eeq

\begin{figure*}
\centering
\includegraphics[width=16cm]{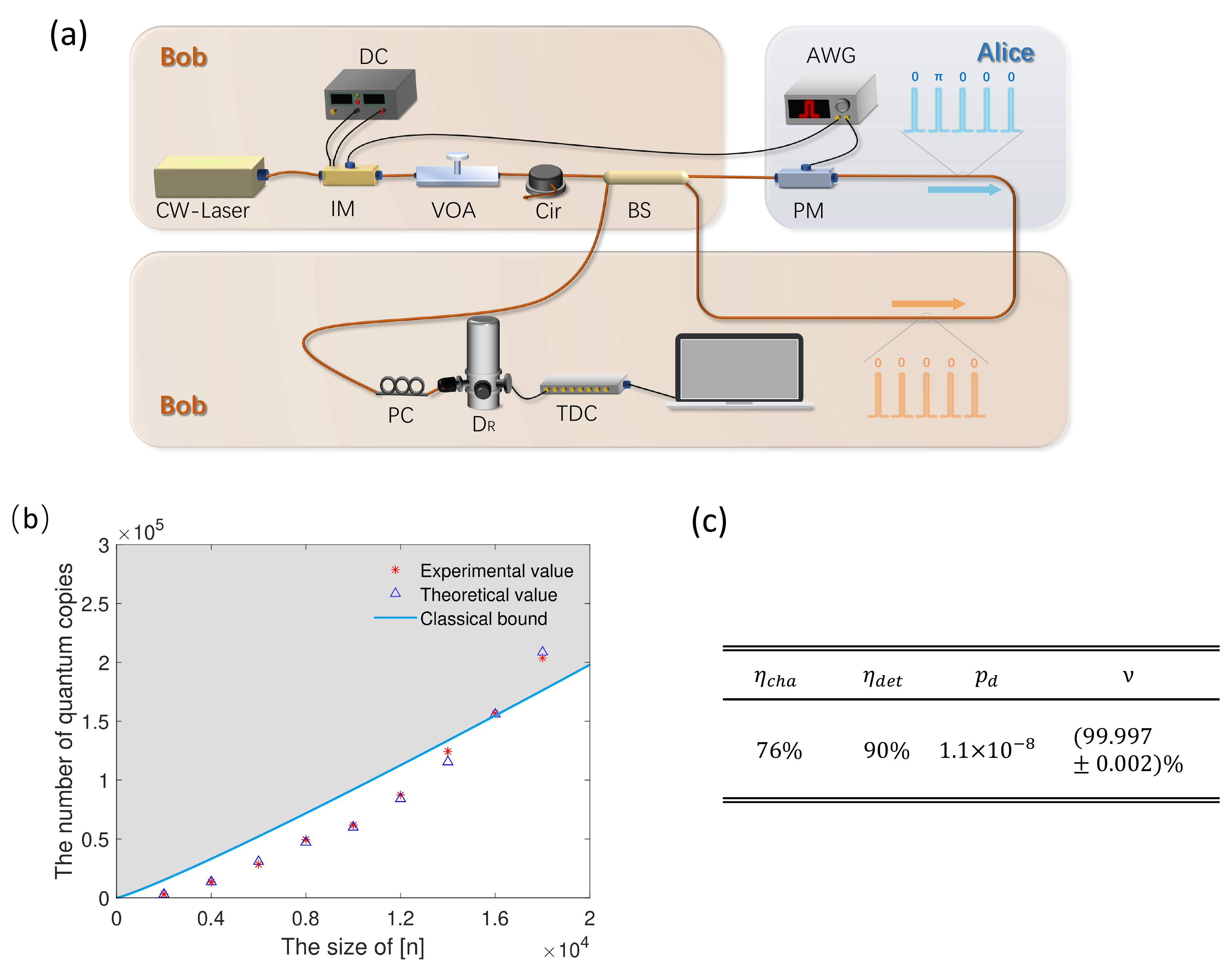}
\caption{(a) Experimental set-up for the coherent-state quantum coupon collector protocol. The optical pulses were generated by a $1550.12$ nm continuous-wave (CW) laser source with an intensity modulator (IM) driven by an external arbitrary waveform generator (AWG). The frequency of the pulse sequences and the duration of a single pulse were 312.5 MHz and 900 ps, respectively. These pulses were attenuated to the single-photon level by a variable optical attenuator (VOA), and then separated into two pulse sequences with different propagation directions by a $50:50$ beam splitter (BS). The clockwise pulse sequence was set to $|\alpha,S\rangle$ after passing a phase modulator (PM) controlled by the AWG. The counterclockwise pulse sequence was used as the local state $|\alpha,[n]\rangle$. Finally, these two pulses interfered at the BS and were detected by a superconducting nanowire single-photon detector $D_R$. A polarisation controller (PC) was used to modify the polarisation of the incident pulses to achieve the maximum detection efficiency. The detection events were recorded using a time-to-digital converter (TDC). (b) Relationship between the required samples and the size of the set $[n]$. We compare the classical lower bound, the samples consumed by our protocol in the practical setup with the experimental parameters listed in Fig~\ref{setup}(c) and its theoretically expected values. The experimental results were in line with the theoretically expected values. For an input size of under $14,000$, our results outperformed the classical lower bound, thus demonstrating quantum advantage. (c) Experimental parameters corresponding to our numerical simulation and experimental demonstration. Note that the channel loss $\eta$ is split into channel efficiency $\eta_{\rm cha}$ and detector efficiency $\eta_{\rm det}$. Here $p_d$ is dark count rate of the single-photon detector and $\nu$ is the limited visibility of the interferometer.}
\label{setup}
\end{figure*}

Based on the above formulae, we present the numerical simulation in Fig.~\ref{protocol}(b-c). The simulation results also guide our experimental demonstration. Figure~\ref{protocol}(b) shows that a higher intensity results in a higher correct probability $P(1,n-1)$. This is because $D_R$ can more easily detect a higher intensity $\left|-\alpha\right\rangle_{i}$ sent by Alice. In contrast, as the visibility $\nu$ of the interferometer is not perfect, higher intensity also increases the probability of $D_R$ clicking when Alice sends $\left|\alpha\right\rangle_{i}$. These two factors lead to the experimental efficiency $E$ and success probability $P_{suc}$ increasing first and then decreasing with an increase in light intensity. $P_{suc}$ directly affects the number of quantum samples required to complete the task. Therefore, we must choose an appropriate light intensity to achieve a balanced trade-off between improving the correct probability and reducing quantum resources.

Figure~\ref{protocol}(c) shows a comparison between the resources required by our protocol and the size of the set $[n]$ with the correct probability $P(1,n-1)=90\%$. To highlight our quantum advantage, we also draw the classical limit of the required samples for comparison. The cost of our protocol is lower than the classical limit when $n<29,000$. When $n<20,000$, the samples required by our protocol are less than half of those required by the classical protocols.

Note that the objective of this article was to construct a specific computational task for experimentally demonstrating quantum advantages in the context of machine learning and communication complexity. Therefore, to simplify the implementation of our protocol, we do not consider the case in which Alice sends incorrect coherent pulses to Bob. In fact, this case belongs to a more complex communication task and is beyond the scope of this article.

\begin{table*}
\center
\caption{Summary of the experimental data. The input size $L$ ranges from $2,000$ to $18,000$ with a step size of $2,000$. For each size $L$, we collected data for $5$ seconds. The table shows the number of coupons sent, the number of detection events within the corresponding time windows, the number of events in which the detector clicked only once in each coupon period, the number of events in which the detector clicked only once and clicks at the correct time bin, the correct probability, the efficiency, the success probability, the minimum number of classical samples required for classical protocols, and the number of quantum samples needed to obtain the correct result utilized in our experiments.}
\begin{tabular}
{c @{\hspace{0.4cm}} c @{\hspace{0.4cm}} c  @{\hspace{0.4cm}} c @{\hspace{0.4cm}} c @{\hspace{0.4cm}} c @{\hspace{0.4cm}} c @{\hspace{0.4cm}} c @{\hspace{0.4cm}} c @{\hspace{0.4cm}}c @{\hspace{0.4cm}} c@{\hspace{0.4cm}}}  \hline \hline
L & $\mu$ & \makecell[c]{Total coupons} & \makecell[c]{Detection \\events} & \makecell[c]{Single \\clicks} & \makecell[c]{Correct \\ clicks} & \makecell[c]{Correct \\probability } & Efficiency & \makecell[c]{Success \\probability } & \makecell[c]{Classical\\ samples} & \makecell[c]{Quantum\\samples}  \\ \hline
2000 & 1 & 781250 & 763766 & 525445 & 490824 & 93.4 \% & 67.2\% & 62.8 \% & $1.52 \times 10^4$ & $3.18 \times 10^3$ \\
4000 & 2 & 390625 & 366475  & 254409 & 233479 & 91.7 \% & 65.1\% & 59.8 \% & $3.32 \times 10^4$ & $1.34 \times 10^4$\\
6000 & 2 & 260416 & 138071 & 120573 & 109603 & 90.9 \% & 46.3\% & 42.1 \% & $5.22 \times 10^4$ & $2.85 \times 10^4$\\
8000 & 3 & 195312 & 135981 & 104831 & 94767 & 90.4 \% & 53.7\% & 48.5 \% & $7.19 \times 10^4$ & $4.95 \times 10^4$\\
10000 & 3 & 156250 & 99214 & 82767 & 75900 & 91.7 \% & 53.0\% & 48.6 \% & $9.21 \times 10^4$ & $6.18 \times 10^4$\\
12000 & 4 & 130208 & 118189 & 79260 & 71477 & 90.1 \% & 60.8\% & 54.9 \% & $1.13 \times 10^5$ & $8.74 \times 10^4$\\
14000 & 5 & 111607 & 96421 & 68674 & 62797 & 91.4 \% & 61.5\% & 56.3 \% & $1.34 \times 10^5$ & $1.24 \times 10^5$\\
16000 & 6 & 97656 & 92380 & 64554 & 59817 & 92.6 \% & 66.1\% & 61.2 \% & $1.55 \times 10^5$ & $1.57 \times 10^5$\\
18000 & 7 & 86805 & 81783 & 57499 & 53706 & 93.4 \% & 66.2\% & 61.9 \% & $1.76 \times 10^5$ & $2.04 \times 10^5$\\
\hline \hline
\end{tabular}

\label{exp_result}
\end{table*}

\bigskip
\noindent
\textbf{Experimental setup and results.} 

We used linear optics components and single-photon detectors to present a proof-of-principle experimental demonstration of the coherent-state quantum coupon collector protocol (Fig.~\ref{setup}(a)). First, the continuous-wave light with a wavelength of $1550.12$ nm emitted by a laser source was carefully modulated into optical pulses at a repetition rate of $312.5$ MHz by using an intensity modulator (IM). To make the modulated waveform as perfect as possible, the modulated light was monitored by a single photon detector instead of an oscilloscope during the modulation phase. Then, the modulated optical pulses were attenuated to the desired level by a variable optical attenuator (VOA) and separated by a $50:50$ BS into two identical pulse sequences $|\alpha,[n]\rangle_A$ and $|\alpha,[n]\rangle_B$. These two pulse sequences travel clockwise and counterclockwise in the Sagnac loop. After passing through a phase modulator (PM), the clockwise pulse sequence $|\alpha,[n]\rangle_A$ is modulated to $|\alpha,S\rangle$ according to the value of the coupon $S$. To convert the counterclockwise sequence $|\alpha,[n]\rangle_B$ into the local state $|\alpha,[n]\rangle$, the PM is turned off when $|\alpha,[n]\rangle_B$ passes through it. We used the Sagnac loop to automatically stabilize the phase fluctuation of the channel to improve interference visibility. In addition, to make the two pulse sequences pass through the PM at different times, one arm of the Sagnac loop was designed to be $1$ m longer than the other arm. The optical circulator was used to prevent the previous optical pulses transmitted back by the BS from affecting the subsequent optical pulses. Finally, the interference results were detected using a superconducting nanowire single-photon detector ($D_R$).

Our scheme combines all copies of the $|S\rangle$ required to learn $S$ into a sequence of $n$ coherent states. Compared with a state that consists of a single photon in $n$ modes, the coherent state sequence is easier to prepare. Moreover, combining all copies of $|S\rangle$ enhances the mean photon number of each time mode of the coherent state sequence, thus increasing the probability of measuring each mode. These features make our scheme easier to implement in experiments, and more effective.

Note that we can only learn $S$ correctly if each element $i \in [n]$ is correctly classified as $S$ or $\bar{S}$. Therefore, even if $p_d$ and $\nu$ have a small effect on a single pulse, it is difficult to correctly classify all elements $i \in [n]$ when $n$ is very large. This means that $p_d$ and $\nu$ limit the maximum size of $[n]$ that our protocol can achieve quantum advantage with a given correct probability. However, in our experiment, because $p_d$ can reach the order of $10^{-8}$ under the current experimental conditions, the effects of the random detection events caused by $p_d$ can be ignored. In contrast, the maximum visibility of an interferometer $\nu$ that can be achieved in a laboratory is on the order of $1-10^{-5}$. As a result, $\nu$ seriously limits the success probability of learning $S$, even if the size of $[n]$ is relatively small. In addition, $\eta$ directly affects the number of quantum samples required to complete the task. Therefore, $\nu$ and $\eta$ are the experimental parameters that need to be improved. When the accuracy of $\nu$ is improved in the future, our protocol can also achieve quantum advantage over $[n]$ with a larger size.

To improve the success probability, we selected a BS whose visibility in the Sagnac loop was nearly $99.9993\%$. We also reduced the magnitude of the dark count probability $p_d$ to $10^{-8}$. At this magnitude, the dark count probability hardly affects the system performance. In addition, we tried to adjust the amplitude of the radio-frequency signal driving the PM to achieve an accurate $\pi$-phase shift.
However, it is almost impossible for a PM to apply a perfect $\pi$-phase shift on a specific pulse without influencing other pulses in a period, especially as the duty cycle in this experimental demonstration is on the order of $10^{-4}\sim 10^{-5}$. Consequently, the number of $D_R$ clicks caused by the pulses $|\alpha\rangle_i$ in a coupon period is much higher than theoretically expected, which means that the experimental interference visibility is lower than that of the BS. Fortunately, during data processing, we found that selecting certain time windows can significantly improve the visibility for interference. However, this approach inevitably filters out certain detection events, thereby affecting the final success probability. By repeatedly selecting different time windows, we achieved a better trade-off between improving visibility and reducing detection events, thereby reducing the number of quantum samples of different coupon lengths.

\begin{figure*}
\centering
\includegraphics[width=18cm]{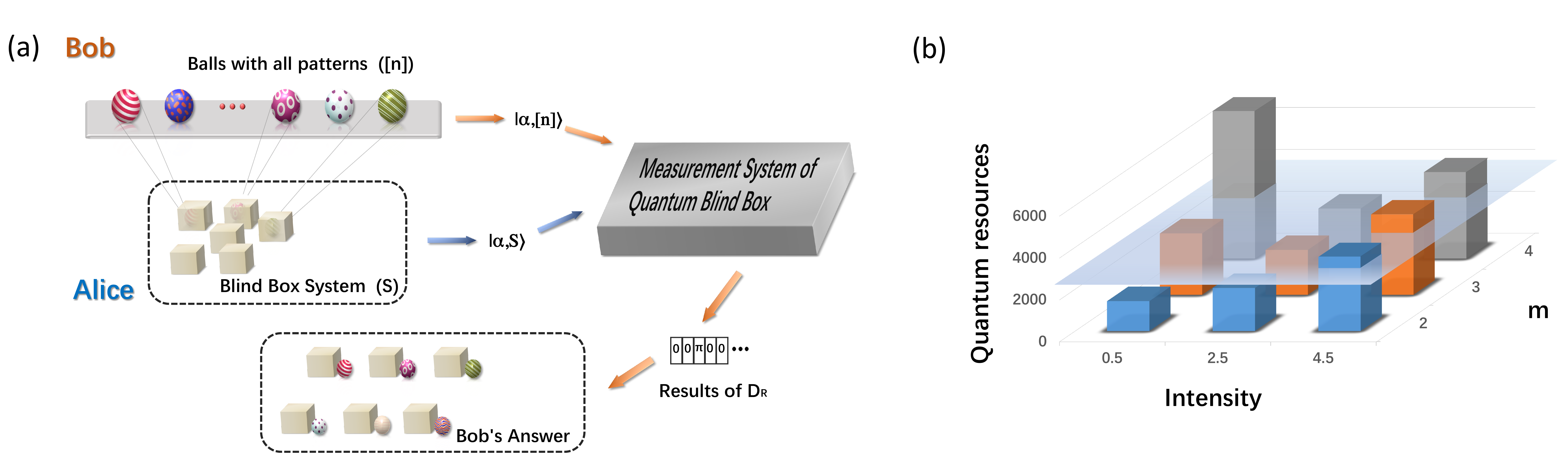}
\caption{(a) Diagram of a quantum blind box. Alice prepares several balls with different patterns ($[n]$). Alice then puts some of these balls into boxes to form a blind box system ($S$) and encodes them into a coherent state $|\alpha,S\rangle$. To determine the balls contained in Alice's blind box system, Bob encodes all balls into a local state $|\alpha,[n]\rangle$. Alice generates $|\alpha,S\rangle$ and $|\alpha,[n]\rangle$ in the measurement system of a quantum blind box, and Bob adjusts their total intensity in the system. Finally, Bob uses the results of $D_R$ to determine which balls are included in Alice's blind box system. (b) Experimental results of the quantum blind box game. For $m$, we tested different values of $m$ with light intensities of $0.5$, $2.5$, and $4.5$. The $Z$-axis represents the expected value of the quantum resources that Bob spends to obtain the correct answer. The plane represents the classical limit. Note that for different $m$, the classical limit is different. Therefore, the plane is not strictly a plane, but it looks like a plane because the gap is too small. The quantum resources at certain light intensities do not exceed the plane, which means that Bob can use these light intensities to break the classical limit.}

\label{blind}
\end{figure*}

The experiment was performed with different sizes of the set $[n]$ ranging from $2000$ to $18000$. For each size $L$, we ran the experiment for $5$ s and analyzed the detection results. The relevant experimental parameters are listed in Fig~\ref{setup}(c). The detailed results are listed in Table~\ref{exp_result}. Our protocol consumes fewer samples than the classic protocol for $L<16000$ (Fig.~\ref{setup}(b)). The gray area in Fig.~\ref{setup}(b) indicates the region in which our protocol consumes more samples than the classical one. By choosing different time windows, we set $\nu = 0.99996$ for $n \leq 6000$ and $\nu = 0.99999$ for $n \geq 8000$. Note that these visibility levels are approximate values based on experimental data. To achieve better experimental results, we did not choose the time window with the most detection events, but the window with the highest visibility. The result is that only a fraction of the photons were detected, which is equivalent to reducing the detection efficiency of detector $D_R$. Therefore, the intensities of the coherent state pulses were modulated much higher than theoretically expected in this experiment. Furthermore, because the quantum resources consumed in our protocol are proportional to the intensity $|\alpha|^2$ of each pulse in $|\alpha, S\rangle$, the degree of photon dissipation directly affects the results of the experiment. When the photon dissipation is larger, a higher intensity is required to compensate for the photon dissipation, thus consuming more quantum resources. Therefore, in our experiments, we strive to improve the channel efficiency $\eta_{cha}$ between Alice and $D_R$ and the detection efficiency $\eta_{det}$ to reduce the consumed quantum resources. Reducing the voltage fluctuations of the phase modulator and selecting a better time window can further improve the experimental results.

\bigskip
\noindent
\textbf{Quantum blind box.} 

Our protocol can not only be used to verify quantum advantage in machine learning from the PAC theory but can also be regarded as a communication task to demonstrate quantum advantage in communication complexity. To this end, we designed a specific application scenario for our protocol, which is called a quantum blind box game. In this game (Fig.~\ref{blind}(a)), Alice acts as a merchant, and Bob acts as a customer. Alice prepares $n$ small balls with different patterns and packs them in boxes to form blind boxes. Alice then takes $n-m$ of these boxes as a blind box system, where $m\geq 1$, and encodes the coherent state $|\alpha,S\rangle$ according to the patterns of the balls in that system. Alice then tells Bob all patterns of the balls and the number of blind boxes in the blind box system.

To determine the patterns contained in the blind box system, Bob uses the same encoding method as Alice to create a local state $|\alpha,[n]\rangle$. The merchant Alice provides the entire measurement system, which is the same as the experimental device shown in Fig.~\ref{setup}(a). Bob can decide the total intensity $\mu$ of the coherent state $|\alpha,S\rangle$ sent by Alice, but the money he needs to pay to Alice is equal to the quantum resources consumed by $|\alpha,S\rangle$. The required quantum resources are $O\left(\mu \log _{2} n\right) $ bits \cite{xu2015experimental,arrazola2014quantum}. Therefore, the higher the intensity $\mu$ of $|\alpha,S\rangle$, the more money Bob needs to pay. Finally, Bob judges the patterns in the blind box system based on the results of the single-photon detector $D_R$. If Bob gives the correct answer, he will be rewarded with $(n-m)\log_2(n-m)\log_2n$ dollars, which is the minimum expected value of the information that needs to be transmitted to obtain the correct answer using classic resources. Therefore, when Bob consumes fewer quantum resources than the classical limit, he can obtain a positive return in this transaction.

\begin{table*}[t]
\caption{Summary of notable quantum advantage demonstrations, with outlines of which fields the demonstrated quantum advantages belong to and the methods used to demonstrate them.}
\begin{ruledtabular}
\begin{tabular}{ccccc}
&\multicolumn{3}{c}{Demonstrated quantum advantages}&\multicolumn{1}{c}{Methods}\\ \hline
& Communication complexity&Machine learning& Computational power  \\
 \hline
This work & YES & YES & - & Linear Optics  \\
Gong \emph{et al.} (2021) \cite{gong2021quantum}& - & - & YES & Superconducting Processor \\
Centrone \emph{et al.} (2021) \cite{centrone2021experimental}& YES & - & - & Linear Optics \\
Zhong \emph{et al.} (2020) \cite{zhong2020quantum} & - & - & YES & Linear Optics \\
Arute \emph{et al.} (2019) \cite{arute2019quantum}&  -  & - & YES & Superconducting processor \\
Kumar \emph{et al.} (2019) \cite{kumar2019experimental}& YES & - & - & Linear Optics \\
Arrazola \emph{et al.} (2018) \cite{arrazola2018quantum}& YES & - & - & Linear Optics \\
Bravyi \emph{et al.} (2018) \cite{bravyi2018quantum}& -  & - & YES & Quantum Circuits\\
Boixo \emph{et al.} (2018) \cite{boixo2018characterizing} & - & - & YES & Quantum Circuits\\
Neville \emph{et al.} (2017) \cite{neville2017classical}& - & - & YES & Linear Optics \\
Xu \emph{et al.} (2015) \cite{xu2015experimental}& YES & - & - & Linear Optics  \\
Bentivegna \emph{et al.} (2015) \cite{bentivegna2015experimental} & - & - & YES & Integrated Photonic Circuits \\
Broome \emph{et al.} (2013) \cite{broome2013photonic}&-& - & YES & Tunable Circuit \\
\end{tabular}
\end{ruledtabular}
\label{compare}
\end{table*}

Let us consider $n = 100$ and $m \in \{2, 3, 4\}$ as an example for demonstrating this game. The experimental results show that for a given value of $m$, the expected quantum resources are significantly affected by the light intensity (Fig.~\ref{blind}(b)). The detailed experimental results are presented in the Methods section. Note that the comparison used herein is the expected value of the quantum resources spent to obtain the correct answer. When the light intensity is small, Bob will not spend a large amount of resources in each game, but it is also difficult to obtain the correct answer. To obtain the bonus, Bob may need to play multiple games. As a result, the expected value of the quantum resources that he spends has increased. This is the reason why quantum resources at a $0.5$ light intensity are relatively high for $m=3$ and $m=4$.

Figure~\ref{blind}(b) also shows that Bob can break the classical limit by choosing the proper intensity for achieving a positive return. In other words, Bob successfully uses quantum resources to design a better strategy than random extraction in this game. This means that in the quantum blind box game, Alice can also allow customers to design various coding methods and measurement strategies for guessing the blind box. The smaller the expected value of the information required for a strategy designed by a customer, the more he can get in return. Overall, quantum advantage in communication complexity has been successfully demonstrated in experiments through the quantum blind box game. We reasonably expect that the ideas contained in this game can be used to design communication protocols with a lower amount of information needed to complete specified communication tasks.

\bigskip
\noindent
\textbf{Discussion}

Prior works have mainly demonstrated the quantum advantages of improving the security of communication and enhancing computational power for specific tasks. Although many studies have attempted to find the superiority of quantum machine learning, these studies have not theoretically proven the quantum advantages of machine learning. In this work, we propose a coherent-state quantum coupon collector protocol and demonstrate it experimentally by using simple linear optical elements and coherent states. Experimental results show that our protocol can effectively reduce the number of samples required to learn coupons exactly with up to $14000$ elements on the basis of a $90\%$ correct probability. Combined with the arguments in Ref. \cite{arunachalam2020quantum}, our result strongly demonstrates the quantum advantages of machine learning under current technology. To compare with quantum advantages achieved by other studies, we summarize them in Table.~\ref{compare}. Note that our scheme does not resort to immature technologies, such as complicated entangled states or ideal single-photon sources. This makes our scheme particularly practical, especially for exemplifying the ability of linear optics.

In addition to the demonstration of the quantum advantage of machine learning based on the PAC learning theory for the first time, we also specifically designed a quantum blind box game based on our protocol and experimentally demonstrated quantum advantage in communication complexity through this game. Our protocol does not save resources exponentially like other communication tasks with quantum advantages. Nevertheless, our protocol can still effectively learn the missing elements in the set $[n]$.  We hope that the ideas contained in this game can inspire other useful applications, such as quantum voting.

Overall, despite potential limitations, our study provides new opportunities for the development of quantum machine learning and quantum communication complexity. We expect that the ability of linear optics can help us achieve more quantum-advantaged communication and computational schemes.

\begin{table*}
    \begin{center}
    \caption{Experimental data of the quantum blind box. We take $n = 100$ as an example and perform phase modulation on $m$ places randomly, where $m \in \{2, 3, 4\}$. For $m \in \{2, 3, 4\}$, the classical resource consumptions are 2985.3, 2948.2, and 2911.2, respectively.}
    \label{tab_blind}
    \begin{tabular}{cc ccc ccc}\hline \hline
      m & intensity & m clicks & \makecell[c]{Correct \\clicks }  &  \makecell[c]{Correct \\probability }  & Efficiency & \makecell[c]{Success \\probability} & \makecell[c]{Quantum \\resources}\\  \hline
      \multirow{3}{*}{2}
       & 0.5 & 119433 & 116560 & 97.6\% & 23.9\% & 23.3\% & 1425.0\\
       & 2.5 & 407768 & 404469 & 99.2\% & 81.6\% & 80.9\% & 2053.3\\
       & 4.5 & 433641 & 431329 & 99.5\% & 86.7\% & 86.2\% & 3465.7\\ \hline
       \multirow{3}{*}{3}
       & 0.5 & 58772 & 56692 & 96.5\% & 11.7\% & 11.3\% & 2929.8\\
       & 2.5 & 392690 & 387474 & 98.7\% & 78.5\% & 77.5\% & 2143.3\\
       & 4.5 & 403131 & 397836 & 98.7\% & 80.6\% & 79.5\% & 3757.5 \\ \hline
       \multirow{3}{*}{4} 
       & 0.5 & 25900 & 24570 & 94.9\% & 5.2\% & 4.9\% & 6760.0\\
       & 2.5 & 353685 & 347275 & 98.2\% & 70.7\% & 69.4\% & 2391.4\\
       & 4.5 & 379154 & 372500 & 98.2\% & 75.8\% & 74.5\% & 4013.1\\
      \hline
    \end{tabular}
  \end{center}
\end{table*}

\bigskip
\noindent\textbf{Materials and Methods}

\noindent\textbf{Selection of time windows.}

As described above, the visibility $\nu$ of the interferometer plays a crucial role in determining the cost in terms of quantum resources. Without phase modulation, $\nu$ in our scheme can easily reach over $99.999\%$. However, imperfections in PM often cause extra counts in unexpected places, which leads to a decrease in visibility.

Fortunately, we find that the visibility varies when different time windows are chosen, which is why the simulation shown in Fig.~\ref{setup}(b) uses two visibilities. The choice of time window also affects the number of detection events. In our experiment, as visibility increased, the number of detected events tended to decrease, which is equivalent to a decrease in the detection efficiency. Therefore, the equivalent detection efficiencies and visibilities for different time windows are different.

To ensure the correct probability $P(m,k)> 90\%$, we traverse different time windows to find the minimum value of the expected value of the required quantum resources. For the expected values displayed shown in Fig.~\ref{setup}(b), we adapted the corresponding experimental parameters according to the time window.

\bigskip
\noindent\textbf{Experimental details of quantum blind box.} 

In the experiment, we correlate the patterns of the balls to the positions of the pulses. The corresponding positions of the balls that are not in the blind box system are loaded with a $\pi$-phase. Therefore, this game can be realized using the experimental setup shown in Fig.~\ref{setup}(a).

The system was run at a repetition rate of $10$ MHz, and each round was $5$ seconds long. The duty cycle of a pulse was approximately $5\%$. The dark count rate per 5 ns detection gate was approximately $6 \times 10^{-7}$. Considering the channel loss, the detection efficiency was approximately $68\%$. The detailed experimental results are presented in Table~\ref{tab_blind}. The experimental apparatus was the same as that used before.

\bigskip
\noindent\textbf{Acknowledgements}\\
We gratefully acknowledge the support from the National Natural Science Foundation of China (No. 61801420), the Natural Science Foundation of Jiangsu Province (No. BK20211145), the Fundamental Research Funds for the Central Universities (No. 020414380182), the Key Research and Development Program of Nanjing Jiangbei New Aera (No. ZDYD20210101), the Key-Area Research and Development Program of Guangdong Province (No. 2020B0303040001) and the China Postdoctoral Science Foundation (No. 2021M691536).


\end{document}